\documentclass[aps,preprint]{revtex4}
\usepackage{bm}
\usepackage{graphicx}
\usepackage{amsmath}
\begin{document}
\bibliographystyle{apsrev}

\title{Low-energy electron scattering in a strong Coulomb field}

\author{A.I.Milstein, I.S.Terekhov}

\affiliation{Budker Institute of Nuclear Physics, 630090
Novosibirsk, Russia}
\date{\today}

\begin{abstract}
The analytic expression for the cross section of low-energy
electron scattering in a strong Coulomb field is obtained. It is
shown that in a wide energy region this cross section differs
essentially from that obtained in the first Born approximation.
\end{abstract}

 \maketitle

\section{Introduction}
The explicit form of the cross section for electron scattering in
a strong Coulomb field at arbitrary energy was obtained many years
ago \cite{Mott29}. This form contains the infinite series with
respect to the angular momentum. Though various approaches to the
summation over the angular momentum were developed in numerous
papers, the numerical calculation of the cross section with the
use of the results of \cite{Mott29} still rather complicated
problem. The detailed review of the papers devoted to the problem
under discussion can be found in \cite{Uberall}. In the papers
\cite{Bartlett40, Doggett56, Sherman56} the numerical calculations
of the cross section were performed for various scattering angles,
the nuclear charge numbers $Z$, and the kinetic electron energy
above $0.023$MeV. For $Z=80$ in the case of backward electron
scattering, it was shown that the ratio of the exact cross section
and the Rutherford cross section increases from $0.15$ to $2.35$
when the electron kinetic energy decreases from $1.675$MeV to
$0.023$MeV. Such big deviation from the non-relativistic result
stimulate the interest to the investigation of the exact  cross
section at very small energy. In the present paper we have solved
this problem calculating the asymptotic form of the cross section
for arbitrary $Z$ and  small kinetic electron energy.

\section{Cross section}
A simple way to derive the scattering amplitude is the use of  the
Green function $G({\bf r}_2,{\bf r}_1|\varepsilon)$   of the Dirac
equation in the external field. The wave function
$\psi_{\lambda{\bf p}}(\bf r)$ can be obtained with the help of
the relation
\begin{equation}\label{Green}
\lim_{ r_1\to \infty }G({\bf r}_2,{\bf
r}_1|\varepsilon)=-\frac{\exp{(ipr_1})}{4\pi
r_1}\sum_{\lambda=1,2}\psi_{\lambda{\bf p}}^{(+)}({\bf
r}_2)\bar{u}_{\lambda{\bf p}}\, ,\quad u_{\lambda{\bf
p}}=\sqrt{\varepsilon+m}
\begin{pmatrix}
\phi_\lambda\\
\dfrac{{\bm \sigma}\cdot {\bm
p}}{\varepsilon+m}\phi_\lambda\end{pmatrix}\, ,
\end{equation}
where $p=\sqrt{\varepsilon^2-m^2}$, $m$ is the electron mass,
$\psi_{\lambda{\bf p}}^{(+)}({\bf r})$ denotes a solution of the
Dirac equation containing at the infinity a plane wave with the
momentum ${\bf p}=-p{\bf n}_1$ (${\bf n}_{1,2}={\bf
r}_{1,2}/r_{1,2}$) and a diverging spherical wave, $\lambda$ is
the helicity, $\hbar=c=1$. In the Coulomb field the
right-hand-side of (\ref{Green}) contains the additional factor
$(2pr_1)^{iq}$, where $q=Z\alpha\varepsilon/p$ , $\alpha=1/137$ is
the fine-structure constant. In \cite{MS82} a convenient integral
representation was obtained for the electron Green function in the
Coulomb field. Using Eqs. (19)-(22) of that paper, we arrive at
the following result:
\begin{eqnarray}\label{wf}
&&\psi_{\lambda{\bf p}}^{(+)}({\bf
r}_2)=\sqrt{\varepsilon+m}\sum_{l=1}^{\infty}
\begin{pmatrix}
f_1\\
\dfrac{p\,{\bm \sigma}\cdot {\bm
n}_2}{\varepsilon+m}f_2\end{pmatrix}\, ,\nonumber\\
&&f_{1,2}=[(R_{1}A+i\frac{mZ\alpha}{p}R_{2}B)M_1 \mp iR_{2}BM_2]\phi_\lambda
\quad ,\nonumber\\
&&A=l\frac{d}{dx}(P_l(x)+P_{l-1}(x))\quad ,\quad
B=\frac{d}{dx}(P_l(x)-P_{l-1}(x))\quad ,\nonumber\\
&& x={\bm n}_1\cdot {\bm n}_2\quad ,\quad      R_{1,2}=1\mp ({\bm
\sigma}\cdot {\bm n}_2)({\bm \sigma}\cdot
{\bm n}_1)\, \nonumber\\
&&M_{1,2}=i\frac{\exp(ipr_2-i\pi\nu)}{pr_2}\int\limits_0^\infty
t^{(\mp 1-2iq)}e^{it^2}J_{2\nu}(2t\sqrt{2pr_2})\,dt \, .
\end{eqnarray}
Here $P_l(x)$ is the Legendre polynomial, $J_{2\nu}$ is the Bessel
function, $\nu=\sqrt{l^2-(Z\alpha)^2}$. The integral in the
expression for $M_{1,2}$ are expressed via the confluent
hypergeometric function. The result (\ref{wf}) is in agreement
with the well-known solution of the Dirac equation in the Coulomb
field.

When $r_2\to\infty$ then the coefficient $W_\lambda$ at the
diverging spherical wave
$$\exp[ipr_2+iq\ln(2pr_2)]/r_2$$ in $\psi_{\lambda{\bf
p}}^{(+)}({\bf r}_2)$ is determined by the asymptotics of the
function $M_1$ in (\ref{wf}) coming  from the region of
integration $t\ll 1$. We have
\begin{eqnarray}\label{ampl}
&&W_{\lambda}=\sqrt{\varepsilon+m}\sum_{l=1}^{\infty}
\begin{pmatrix}
f\\
\dfrac{p\,{\bm \sigma}\cdot {\bm
n}_2}{\varepsilon+m}f\end{pmatrix}\, ,\nonumber\\
&&f=\frac{ie^{-i\pi\nu}\Gamma(\nu-iq)}{2p\Gamma(\nu+1+iq)}
[R_{1}A+i\frac{mZ\alpha}{p}R_{2}B]\phi_\lambda \quad .
\end{eqnarray}
As a result we  obtain the cross section
\begin{eqnarray}\label{cross}
&&\frac{d\sigma}{d\Omega}=\frac{2}{p^2}\left\{(1+x)|F^{\prime}|^2
+\left(\frac{mZ\alpha}{p}\right)^2\frac{|F|^2}{1-x}\right\}
\quad ,\nonumber \\
&&F=-\frac{i}{2}\sum_{l=1}^{\infty}le^{i\pi(l-\nu)}\frac{\Gamma(\nu-iq)}
{\Gamma(\nu+1+iq)}\,[P_l(x)-P_{l-1}(x)]\quad , \quad
x=\cos\vartheta \, .
\end{eqnarray}
that coincides with the Mott result \cite{Mott29}.

\section{Low-energy scattering}

\begin{figure}[htb]
\centering
\includegraphics*[width=0.7\textwidth]{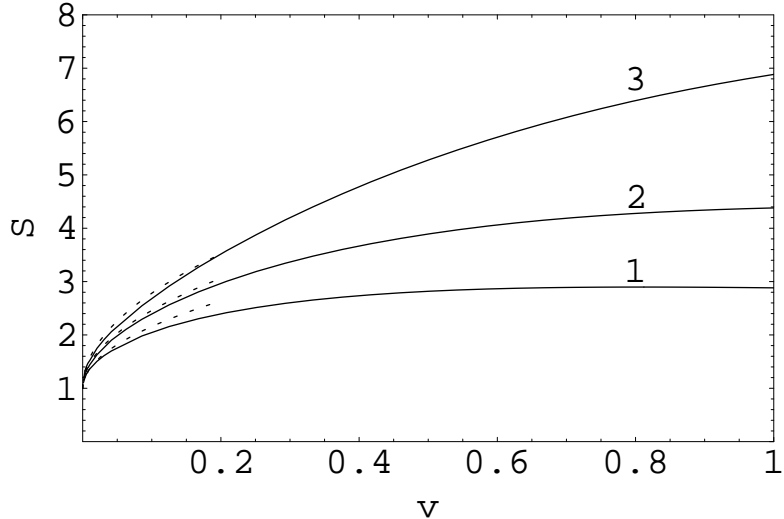}
\caption{The dependence of S on $v$ for $x=-1$ and $Z\alpha=0.6 $
(1), $0.7$ (2), and $0.8$ (3). Solid lines are the exact results,
the dashed lines are the asymptotics. } \label{fig1}
\end{figure}

Let us consider now the cross section (\ref{cross}) in the limit
$q=Z\alpha\varepsilon/p\gg 1$ , $Z\alpha\sim 1$  corresponding to
the low-energy electron scattering in the strong Coulomb field. In
the first Born approximation the cross section reads
\cite{Mott29}:
\begin{equation} \frac{d\sigma_{B}}{d\Omega}=\frac{q^2}{p^2(1-x)^2}
[1-\frac{p^2}{2\varepsilon^2}(1-x)]\quad .
\end{equation}
Using the asymptotics of the $\Gamma$-function in (\ref{cross}) we
represent the ratio $S=d\sigma/d\sigma_{B}$ in the  following
form:
\begin{eqnarray}\label{crossasymp1}
&&S=1+\frac{(1-x)}{q}\mbox{Im}\Biggl\{
\exp\left[-iq\ln\left(\frac{1-x}{2}\right)\,\right]\nonumber\\
&&\times\sum_{l=1}^{\infty}l(-1)^l[P_l(x)-P_{l-1}(x)]
\left(e^{-2i\pi\nu}-1\right)e^{il^2/q}\Biggr\}\, .
\end{eqnarray}
If $(1+x)\gg 1/q$ then it is possible to neglect the factor
$\exp(il^2/q)$ in  (\ref{crossasymp1}). We obtain
\begin{eqnarray}\label{crossasymp1Q}
&&S=1+\frac{(1-x)}{q}\mbox{Re}\Biggl\{\exp\left[-iq\ln\left(\frac{1-x}{2}\right)\,\right]
\nonumber\\
&&\times \Biggl\{
\pi(Z\alpha)^2\left(\sqrt{\frac{2}{1+x}}-1\right)
-i\pi^2(Z\alpha)^4\ln\sqrt{\frac{1+x}{2}}\nonumber\\
&& -i\sum_{l=1}^{\infty}l(-1)^l[P_l(x)-P_{l-1}(x)]
\left[e^{-2i\pi\nu}-1-\frac{i\pi(Z\alpha)^2}{l}+\frac{\pi^2(Z\alpha)^4}{2l^2}
\right]\Biggr\}\Biggr\}\quad .
\end{eqnarray}
Thus, at $(1+x)\gg 1/q$ the correction to the cross section
$\propto 1/q$. Note that the sum in (\ref{crossasymp1Q}) converges
very rapidly for any $x$. When $(1+x)\sim 1/q$ (backward
scattering), then the main contribution to the sum in
(\ref{crossasymp1}) comes from $l\sim \sqrt{q}\gg 1$. Using the
asymptotics of the Legendre polynomials at $x\to -1$ and replacing
the summation by the integration we get
\begin{eqnarray}\label{crossasymp2Q}
S=1+(1-x)\frac{\pi^{3/2}(Z\alpha)^2}{\sqrt{q}}
\cos\left[\frac{\pi}{4}+\frac{q(1+x)}{4}\,\right]
J_0\left(\frac{q(1+x)}{4}\right)\quad .
\end{eqnarray}
We see that at $(1+x)\sim 1/q$ the correction to the cross section
$\propto 1/\sqrt{q}$. Therefore, for the low-energy scattering the
biggest difference between the exact in $Z\alpha$ cross section
and $d\sigma_B/d\Omega$ is in the case of backward scattering. For
arbitrary $q$ and $x=-1$ the function  $S$ reads (see
(\ref{cross})):
\begin{eqnarray}\label{crossback}
&&S=4\left|\sum_{l=1}^{\infty}le^{-i\pi\nu}\frac{\Gamma(\nu-iq)}
{\Gamma(\nu+1+iq)}\right|^2 \, .
\end{eqnarray}
In Fig.~\ref{fig1} we show the dependence of $S$ on
$v=p/\varepsilon$ for $x=-1$ and various values of $Z$, and
compare it with the low-energy asymptotics
$S=1+\sqrt{2}\,\pi^{3/2}(Z\alpha)^{3/2}\sqrt{v}$.

We see that for backward scattering the exact in $Z\alpha$ cross
section differs essentially from $d\sigma_B/d\Omega$ up to rather
small $v$ , and this difference decreases very slowly ($\propto
\sqrt{v}$). Strictly speaking, the asymptotics are valid when
$(S-1)\ll 1$. Nevertheless, it is seen from Fig.\ref{fig1} that
the difference between the exact result and the asymptotics is
small starting from $v< 0.2$. For $(1+x)\gg 1/q$ one can check
that $(S-1)\ll 1$ starting from $q>10$.

This work was supported through Grant RFBR  01-02-16926.

\end{document}